\documentclass[twocolumn,multicolumn,aps,prb,showpacs]{revtex4}
\usepackage{graphicx}
\usepackage{dcolumn}
\usepackage{bm}
\usepackage{color}
\usepackage{latexsym}
\usepackage[T1]{fontenc}
\usepackage{amsmath}

\input{epsf}

\begin{document}

\preprint{APS/123-QED}

\date{\today}

\title{
Flux pinning in (1111) iron-pnictide superconducting crystals}
\author{C.J. van der Beek$^{1}$, G. Rizza$^{1}$, M. Konczykowski$^{1}$, 
P. Fertey$^{2}$, I. Monnet$^{3}$, Thierry Klein$^{4}$, R. Okazaki$^{5}$, 
M. Ishikado$^{6}$, H. Kito$^{7,8}$, A. Iyo$^{7,8}$, H. 
Eisaki$^{7,8}$, S. Shamoto$^{6,8}$, M.E. Tillman$^{9}$, S.L. Bud'ko$^{9}$, P.C. Canfield$^{9}$, T. Shibauchi$^{5}$, and Y. Matsuda$^{5}$}
\affiliation{$^{1}$Laboratoire des Solides Irradi\'{e}s, CNRS UMR 7642 \& CEA-DSM-IRAMIS, Ecole Polytechnique, 91128 Palaiseau, France \\
$^{2}$Synchrotron SOLEIL, L'Orme des Merisiers, Saint-Aubin - BP 48, F-91192 Gif-sur-Yvette cedex, France \\
$^{3}$CIMAP, 6 boulevard du Mar\'{e}chal Juin, F-14050 Caen cedex 4, France \\
$^{4}$Laboratoire Louis N\'{e}el, CNRS-UPR 5031, F-38042 Grenoble cedex, France\\
$^{5}$Department of Physics, Kyoto University, Kyoto 606-8502, Japan \\
$^{6}$Quantum Beam Science Directorate, Japan Atomic Energy Agency, Tokai, Naka, Ibaraki 319-1195, Japan \\
$^{7}$Nanoelectronics Research Institute (NeRI), National Institute of Advanced Industrial Science and Technology (AIST), 1-1-1 Central 2, Tsukuba, Ibaraki 305-8568, Japan\\
$^{8}$ JST, Transformative Research-Project on Iron Pnictides (TRIP), Chiyoda-ku, Tokyo 102-0075\\
$^{9}$Department of Physics \& Astronomy and Ames Laboratory, Iowa State University, Ames, Iowa, U.S.A}

\begin{abstract}
Local magnetic measurements are used to quantitatively characterize 
heterogeneity and flux line pinning in  
PrFeAsO$_{1-y}$ and NdFeAs(O,F) superconducting single crystals. In spite of 
spatial fluctuations of the critical current 
density on the macroscopic scale, it is shown that the major contribution 
comes from collective pinning of vortex lines by microscopic defects by the 
mean-free path fluctuation mechanism. The defect density extracted 
from experiment corresponds  to the dopant atom density, which means 
that dopant atoms 
play an important role both in vortex pinning and in 
quasiparticle scattering. In the studied underdoped PrFeAsO$_{1-y}$ and 
NdFeAs(O,F) crystals, there is a background of strong pinning, which we attribute to 
spatial variations of the dopant atom density on the scale of a few 
dozen to one hundred nm.  These variations do not go beyond 5 \% -- we 
therefore do not find any evidence for coexistence of the 
superconducting and the antiferromagnetic phase. The critical current 
density in sub-T fields is characterized by the presence of a peak 
effect, the location of which in the $(B,T)$--plane is consistent 
with an order-disorder transition of the vortex lattice. 
\end{abstract}

\maketitle

\section{Introduction}

The characterization of the physical properties of new superconducting 
materials such as the recently discovered iron-pncitide 
superconductors \cite{Kamihara2006,Kamihara2008,Takahashi2008,Chen2008,Ren2008,Ren2008ii,Kito2008} requires a good 
knowledge of sample morphology and microstructure. The measurement 
and interpretation of thermodynamic quantities such as the magnetization, 
the magnetic torque,\cite{Kubota2008} or the specific heat, or transport properties 
such as the resistance or irreversible magnetization, may be complicated by 
material inhomogeneity on mesoscopic or macroscopic length scales. On the 
other hand, microscopic disorder is well-known to be beneficial for 
vortex line pinning and high critical currents. Finally, from the 
defect-vortex interaction, one might hope to extract information on 
electronic scattering mechanisms in the iron-pnictide 
superconductors, as well as on the premise of phase co-existence. In 
underdoped pnictides especially, it has been argued that the 
coexistence of the low-doping 
anti-ferromagnetic state and the superconducting state at higher doping levels may affect physical 
properties.\cite{Drew2008}


Vortex pinning and the critical current density in the iron pnictide 
superconductors has mainly focussed on the so-called ``122'' 
compounds, since large single crystals of these are available. Most 
notably, magnetic flux penetration in 
Ba(Fe$_{0.93}$Co$_{0.07}$)$_{2}$As$_{2}$ has been studied using 
magneto-optical imaging by Prozorov {\em 
et al.}.\cite{Prozorov2008,Prozorov2009} The same authors reported on the 
irreversible magnetization and flux creep in this compound, and 
found qualitative agreement with collective creep in the so-called bundle 
regime.\cite{Blatter94} The non-monotonous behavior of the sustainable current as 
function of magnetic field was interpreted in terms of a crossover to plastic 
creep.\cite{Prozorov2008} A similar behavior was found for crystals 
with different doping levels;\cite{Prozorov2009,Shen2010} the overall 
behavior of the critical current density as function of doping was attributed 
to the changing density of structural domain walls, that act as 
strong pinning centers.\cite{Prozorov2009ii} Yamamoto {\em et al.} 
obtained similar results on the same Ba(Fe$_{0.9}$Co$_{0.1}$)$_{2}$As$_{2}$, but attributed the temperature- and 
field-dependent features of the critical current density to an 
inhomogeneous distibution of Co atoms.\cite{Yamamoto2009} 
Very large critical currents, as well as a non-monotonous width of the 
irreversible magnetization loops correponding to a peak-effect in the critical 
current\cite{Kokkaliaris99,Rassau99,vdBeek2000ii,Paltiel,Klein2010} were measured by Yang {\em et al.} in single 
crystalline Ba$_{0.6}$K$_{0.4}$Fe$_{2}$As$_{2}$,\cite{Yang2008ii} who 
concluded to the presence of small-sized normal state regions in their 
samples. Finally, irreversible magnetization and flux creep 
measurements were conducted on SmFeAsO$_{0.9}$F$_{0.1}$ \cite{Yang2008} 
and polycrystalline NdFeAsO$_{0.82}$Fe$_{0.18}$,\cite{Prozorov2008ii,Wang2008} 
members of the ``1111'' family of compounds. In all the above cases, 
the critical current at low fields was characterized by a peak and 
negligible magnetic relaxation, followed by more pronounced thermally 
activated flux motion at higher fields, which was found to be in qualitative 
agreement with the collective creep theory.\cite{Blatter94} However, 
no quantitative analysis of the data has been performed, and no 
definite consensus as to the defects at the origin of flux pinning 
has been established.


The aim of the present paper is 
the identification of defects responsible for flux pinning 
in single crystals of the (Re)FeAsO ``1111'' family of superconducting compounds. 
The microstructure is characterized by the undulation of the FeAs layers and 
the presence of sparse nanometer-sized defects, both of which do not seem to influence flux pinning. 
The largest contribution to the critical current $j_{c}$ is shown to 
arise from the dopant atoms, which act as scatterers for 
quasi-particles in the vortex cores. One therefore deals with pinning 
by local variations of the mean-free path ($\delta \kappa$ mechanism). 
The temperature- and field dependence of $j_{c}$ is 
very well described by collective flux pinning in the 
single-vortex limit, but superposed on a strong pinning contribution 
arising from small fluctuations of the doping level on the scale of 
dozens of nm.

\section{Experimental Details}

PrFeAsO$_{1-y}$ crystals (with the P4/mmm structure) were grown at 
1300$^{\circ}$C 
and 2 GPa from pressed pellets consisting of the starting materials 
PrAs, Fe, and Fe$_{2}$O$_{3}$, in the nominal composition 
PrFeAsO$_{0.6}$.\cite{Ishikado2009,Okazaki2009}  The typical size of the crystals is $100 \times 100 
\times 30$ $\mu$m$^{3}$; the average final composition corresponds to $y \sim 0.1$. 
A number of monolithic crystals from this batch has been previously 
used for the measurement of the superfluid density,\cite{Hashimoto2009}  the field of first flux 
penetration,\cite{Okazaki2009} and the electrical resistivity in the vicinity of 
the upper critical field $B_{c2} = \Phi_{0}/2\pi\xi^{2}$ 
(with $\xi$ the coherence length).\cite{Okazaki2009} The 
superconducting properties of the compound are therefore completely 
characterized. The temperature dependence of the in-plane penetration depth 
$\lambda_{ab}(T)$ (for currents parallel to the $ab$ plane) is well described by a simple two-gap model, without 
any nodes of the order parameter. The magnitude of the low temperature 
penetration depth is $\lambda_{ab}(0) = 280$ nm.\cite{Okazaki2009} 
Table~\ref{table:parameters} gathers the superconducting parameters of 
PrFeAsO$_{1-y}$, including the characteristic energy $\varepsilon_{0} 
\equiv \Phi_{0}^{2}/4\pi\mu_{0}\lambda_{ab}^{2}$ (corresponding to 
$4\xi^{2}$ times the condensation energy), the low-field anisotropy 
ratio $\varepsilon_{\lambda} \equiv \lambda_{ab}/\lambda_{c}$, and the depairing 
current $j_{0} \equiv 4 \varepsilon_{0} / \sqrt{3} \Phi_{0} \xi$ 
($\mu_{0} \equiv 4\pi\times 
10^{-7}$~Hm$^{-1}$).

\begin{figure}[t]
\includegraphics[width=0.3\textwidth]{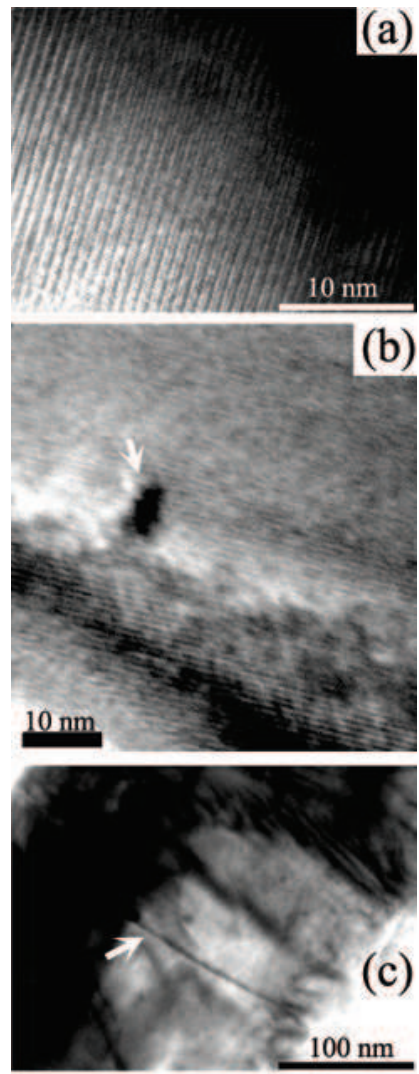}
\caption{Transmission electron microscopy images of single crystalline 
PrFeAsO$_{1-y}$ (a) High resolution bright field image, revealing the
 undulation of the FeAs layers. (b) Bright field micrograph of a zone 
 containing a nm-size inclusion (indicated by the white arrow). (c) Bright field image revealing 
 contrast due to a line dislocation (indicated by the arrow).}
\label{fig:TEM}
\end{figure}

Several PrFeAsO$_{1-y}$ crystals were prepared for Transmission 
Electron Microscopy (TEM). In each case, two crystals, of lateral 
dimensions $\sim 100$ $\mu$m, were glued between 0.5 mm thick Si 
platelets; these were then thinned down until the crystals were flush 
with the edges. Further thinning yielded sections parallel to the 
$c$-axis, suitable for TEM. Figure~\ref{fig:TEM}a, a high 
resolution image of one of the sections, shows clear contrast 
corresponding to the FeAs planes, with some undulation. The 
presence of 5 - 10 nm sized defects, possibly secondary phase 
precipitates, is also observed (Fig.~\ref{fig:TEM}b). These defects 
are separated by a distance of the order of several dozen to several 
hundred nm, depending on location. Finally, Fig.~\ref{fig:TEM} shows 
contrast associated with the presence of a linear dislocation core, 
occasional examples of which were found.

\begin{figure}[tb]
\includegraphics[width=0.45\textwidth]{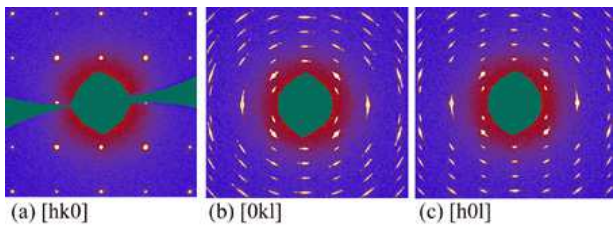}
\caption{(color online) Cuts through reciprocal space of the PrFeAsO$_{1-y}$ 
compound, reconstructed from synchrotron radiation X-ray diffraction 
on crystal \# 7 (the same crystal as in 
Ref.~\protect\cite{Okazaki2009}). (a) the [hk0] plane (b) the [0kl] 
plane; (c) the [h0l] plane.}
\label{fig:SOLEIL}
\end{figure}

Three PrFeAsO$_{1-y}$ crystals (\# 1, 3, and 7) were characterized by X-ray 
diffraction using 28.3 keV (0.43811 \AA) radiation on the CRISTAL beam at the SOLEIL 
synchrotron. Images of diffraction spots were collected on a 2D CCD detector when the 
sample was rotated around an axis in the $ab$ plane. From the 360 measured 
images,  successively collected during 1 s after a progression of 1$^{\circ}$ of the 
crystal rotation, layers of reciprocal 
space were numerically reconstructed. Three such sections, containing 
the origin, are shown in Fig.~\ref{fig:SOLEIL}. The [hk0] section 
reveals very good translational order in the basal plane. However, 
the fulfilment of the Laue condition over extended streaks in the [h0l] 
and [0kl] planes shows that crystalline order along the $c$-axis is not 
as good. The pronounced elongation of both low- and higher order nodes 
in the [00l] direction indicates that this disorder more than likely 
originates from the undulation of the planes 
observed in TEM. Other kinds of $c$-axis disorder such as stacking 
faults or anti-phase boundaries would have yielded a larger 
broadening of nodes outside the [hk0] plane, as compared to the lower 
order nodes. From the elongation of the nodes at [100] and [010], we 
estimate the buckling of the layers to result in a variation of their 
orientation of up to 5$^{\circ}$. The same results were obtained for 
all studied crystals.

The NdFeAsO$_{0.9}$F$_{0.1}$ crystals used in this study are the same as 
that of Ref.~\onlinecite{Kacmarcik2009}; they were synthesized at high pressure 
in a cubic, multi-anvil apparatus. The crystals, extracted from a  
polycrystalline batch, had dimensions $210 \times 
320 \times 30$ $\mu$m$^{3}$ (\# 1) and $150 \times 200 \times 50$ 
$\mu$m$^{3}$ (\# 2), and critical 
temperatures $T_{c} = 34.5 \pm 1.5$ K (\# 1) and $37.5 \pm 1$ K (\# 2). 
The superconducting parameters of NdFeAs(O,F) of this particular 
doping level have been studied in Refs.~\onlinecite{Kacmarcik2009} 
and \onlinecite{Pribulova2009}, and are summarized in Table~\ref{table:parameters}.

\begin{table}[b]
    \begin{tabular}{lcccccc}
	\hline
	\hline
	compound & $\lambda_{ab}(0)$ & $\varepsilon_{\lambda}(0) $ & 	
	$\xi_{ab}(0)$ & $\varepsilon_{0}(0)$ & $j_{0}(0)$ & $Gi$ \\
        \hline
	PrFeAsO$_{0.9}$           \protect\cite{Okazaki2009}                 &   280 nm &  0.4  & 1.8 nm  & $3.2 \times 10^{-12}$ Jm$^{-1}$ & $2 \times 10^{12}$ Am$^{-2}$ & $3 \times 10^{-3}$ \\
	NdFeAsO$_{0.9}$F$_{0.1}$ \protect\cite{Pribulova2009,Kacmarcik2009} & 	
	$270\pm 40$ nm & 0.25  & 2.4 nm  & $3.5 \times 10^{-12}$ Jm$^{-1}$ & 
	$1.6 \times 10^{12}$  Am$^{-2}$ & $3 \times 10^{-3}$ \\
	\hline
	\hline
    \end{tabular}
    \caption{Superconducting properties of the crystals used in this study.}
    \label{table:parameters}
    \end{table}

In order to obtain the value and local distribution of $T_{c}$ and $j_{c}$, 
flux penetration into the superconducting crystals was imaged using 
the direct magneto-optical imaging (MOI) method.\cite{Dorosinskii92} 
Crystalline inhomogeneity in the vicinity of the critical temperature 
was characterized using the Differential Magneto-Optical (DMO) method.\cite{Soibel99} In 
MOI, a ferrimagnetic garnet indicator with in-plane anisotropy is 
placed on top of the sample under study, and observed using a 
polarized light microscope. The presence of a non-zero perpendicular 
component $B_{\perp}$ of the magnetic induction is revealed, by 
virtue of the Faraday effect of the garnet, as a non-zero intensity of 
reflected light when the polarizers of the microscope are (nearly) 
crossed. Thus, light areas in the MO images correspond to areas of 
high perpendicular induction, while dark regions have small or zero 
$B_{\perp}$. In DMO, magneto-optical images taken at applied fields 
$H_{a}$ and $H_{a}+\Delta H_{a}$ (with $\Delta H_{a} = 1$ Oe) are 
subtracted; the procedure is repeated 100 times, and the subtracted 
images averaged.


The local critical current density of the investigated crystals was 
obtained by calibrating the luminous intensity of the MOI images, so 
as to obtain a map of the local induction. $j_{c}$ was then determined  
as twice the  gradient of the local flux density, measured over an 
interval of length 20 $\mu$m  
perpendicular to the sample boundary, 
and averaged over a width of 20 $\mu$m, 
(parallel to the sample boundary).  This procedure is justified in that, given our crystals'  
aspect ratio, flux profiles at the sample surface are nearly 
linear.\cite{Brandt96} In what follows, the area over which $j_{c}$ was measured 
was chosen such that $\langle B \rangle \approx 300$ Oe over the 
$20\times 20$ 
$\mu$m$^{2}$ region.

Further measurements were carried out using a micron-sized Hall probe 
array, tailored in pseudomorphic GaAlAs/GaAs heterostructure. 
\cite{Shibauchi2007,Okazaki2009} The 10 Hall sensors of the array had an active area 
of $3 \times 3$ $\mu$m$^{2}$, while an eleventh sensor was used for the 
measurement of the 
applied field. The Hall probe magnetometry technique is complementary to 
magneto-optical imaging in that it has greater sensitivity and 
can be used up to substantially higher magnetic fields; on the other 
hand, it only allows the measurement of $B_{\perp}$ along the array of 
sensors, and not over the entire two-dimensional sample surface.

\begin{figure}[t]
\includegraphics[width=0.22\textwidth]{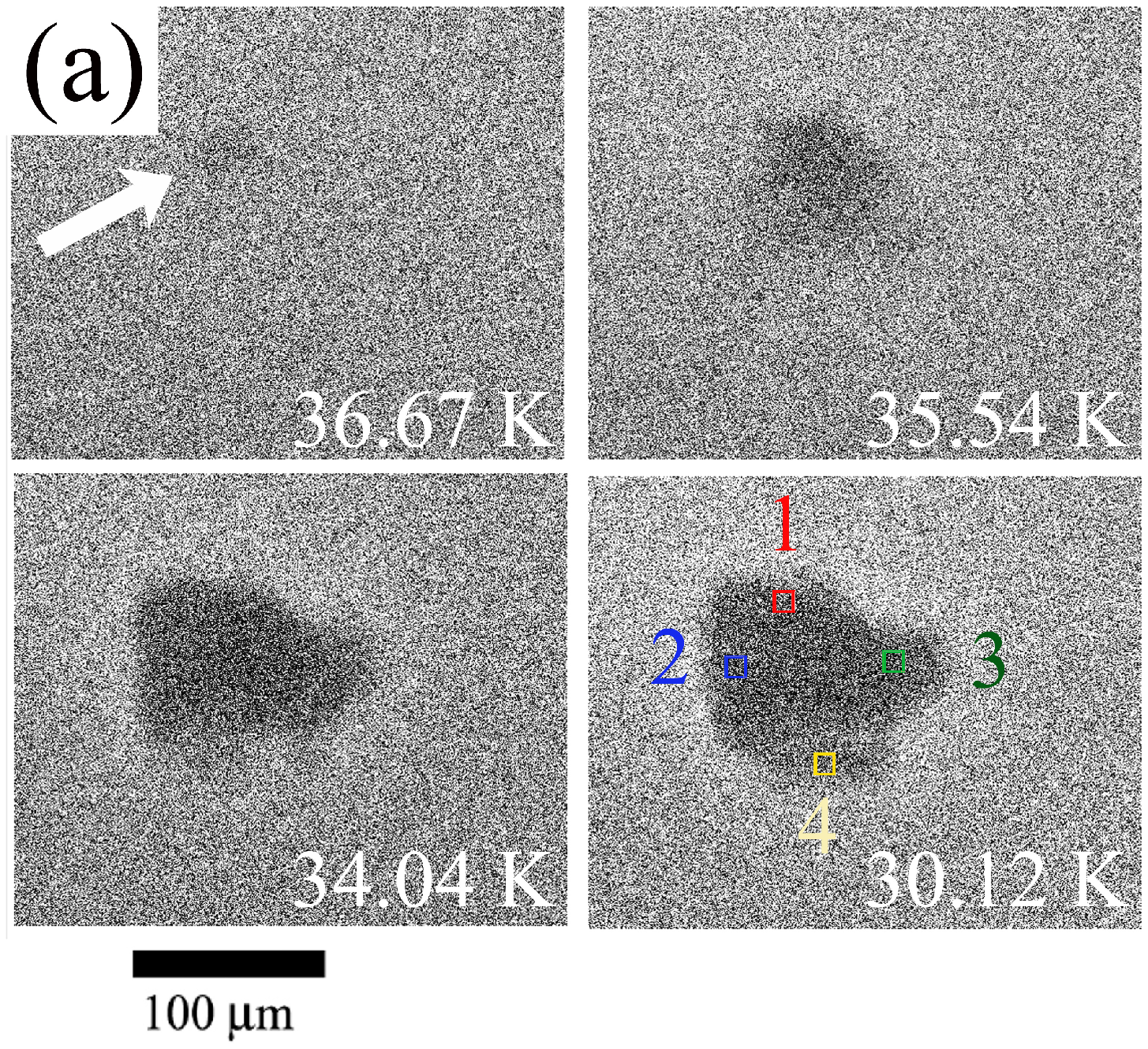}
\includegraphics[width=0.24\textwidth]{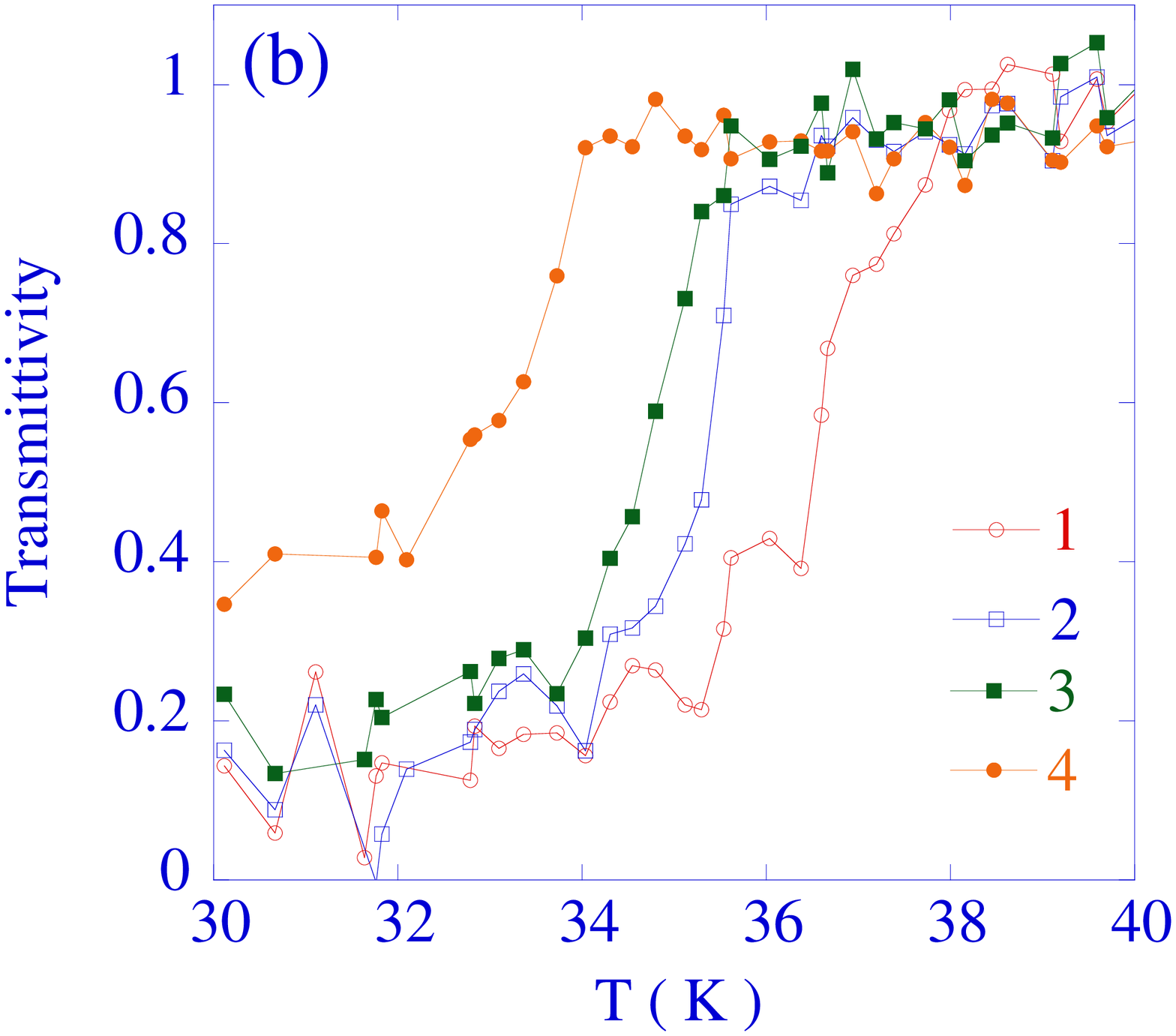}
\caption{(color online) (a) DMO images of the screening of an ac field of 1 Oe by 
PrFeAsO$_{1-y}$ (bi)crystal \# 1, at various temperatures close to $T_{c}$. 
Screening first appears in the upper left-hand corner by the 
crystal (indicated by the arrow). The crystal progressively transits 
to the superconducting state between 38 and 32 K. (b) Local 
permeability, defined as the ratio $(I(\mathbf r, T) - 
I(\mathbf r, T\ll T_{c})/(I(\mathbf r,T\gg T_{c} - I (\mathbf r,T\ll 
T_{c})$ of the local luminous intensities $I(\mathbf r,T)$, in the 
zones 1--4 indicated in (a).}
\label{fig:inhomogeneous}
\end{figure}

\newpage

\section{ Results}
\label{section:results}

\subsection{PrFeAsO$_{1-y}$ }


Spatial inhomogeneity of the critical temperature 
in single crystals was investigated by the DMO images near the transition. 
A typical example is shown in 
Fig.~\ref{fig:inhomogeneous}a, depicting four DMO images, acquired 
with a $\Delta H_{a} = 1$ Oe modulation in the absence of a static 
field, at various temperatures spanning the normal-to-superconducting 
transition. In this particular case, diamagetic screening first 
appears at $T \sim 38$ K in the upper left-hand corner of the 
crystal. Magnetic flux is progressively excluded from the crystal 
bulk, until the largest part is fully screened at $T = 34$ K. 
However, the small grain at the bottom is only  fully screening at $ 
T = 31$ K. Fig.~\ref{fig:inhomogeneous}b shows the ac permeability, 
determined from the luminous intensities $I(\mathbf r,T)$ as $T_{MO} = (I(\mathbf r, T) - 
I(\mathbf r, T\ll T_{c})/(I(\mathbf r,T\gg T_{c} - I (\mathbf r,T\ll 
T_{c})$, for four regions indicated in the last panel of 
Fig.~\ref{fig:inhomogeneous}a. It is seen that, locally, the crystal 
shows sharp transitions to the superconducting state. However, a 
global measurement (\em e.g. \rm by a commercial magnetometer) would 
clearly result in a 
broadened transition.

\begin{figure}[tb]
\includegraphics[width=0.4\textwidth]{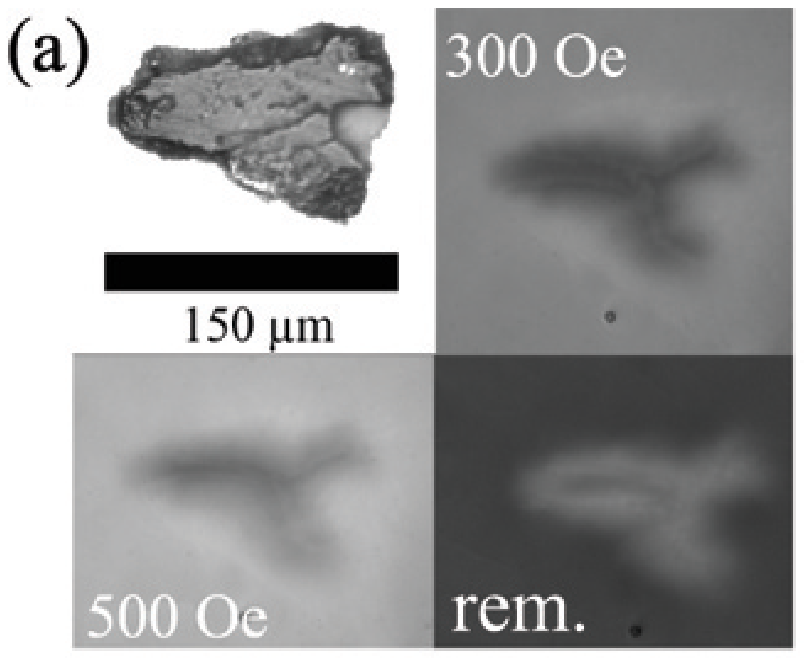}
\includegraphics[width=0.4\textwidth]{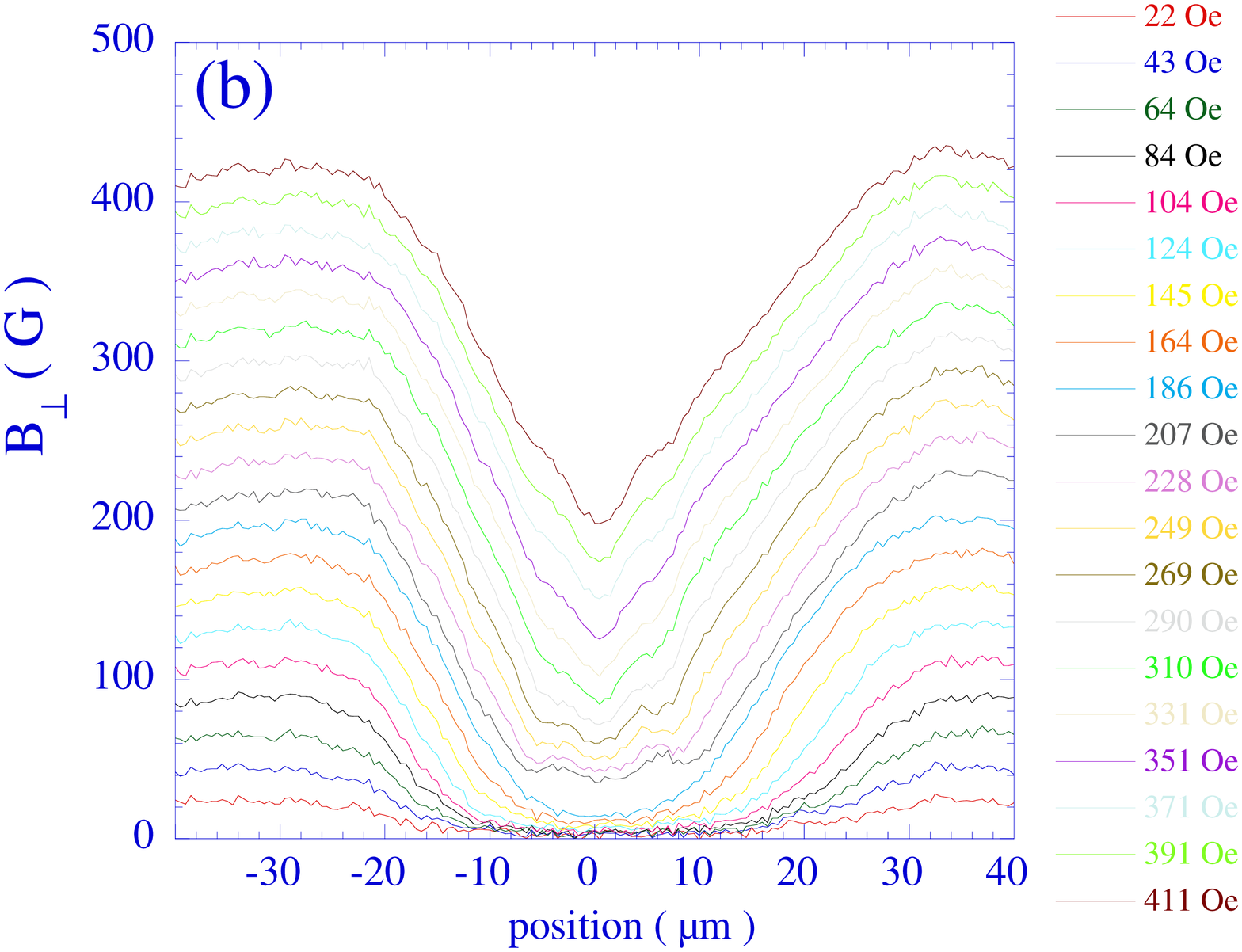}
\caption{(color online) (a) Direct MOI of the screening of an dc field by 
PrFeAsO$_{1-y}$ crystal \# 7 (the same crystal as in 
Ref.~\protect\cite{Okazaki2009}), at 7.1 K. Shown are a polarized 
light image of the crystal; and MOI for $H_{a} = 300$ Oe, $500$ Oe, 
and the trapped flux in zero field (after the application of 500 Oe).  
(b) Flux profiles, measured from top to bottom across 
the central part of this crystal, for increasing values of the applied 
magnetic field $H_{a}$, and a temperature of 11 K.}
\label{fig:Bean}
\end{figure}

Local values of the critical current density $j_{c}$ are obtained from the MO imaging of the largest grains in polycrystalline 
conglomerates, or from the flux distribution in monolithic crystals, such as depicted in 
\ref{fig:Bean}a, for PrFeAsO$_{1-y}$ crystal \# 7. The magnetic flux 
distributions in such crystals are characteristic of the Bean critical 
state;\cite{Brandt96,Bean62,Brandt93,Zeldov94} Fig.~\ref{fig:Bean} 
shows an example of  profiles  obtained across the central part of crystal \# 7 at $T = 11$ K. Due to 
the relatively large thickness-to-width ratio of the crystal, $d/w 
\sim 0.3$, flux profiles resemble straight lines; $j_{c} \approx \frac{1}{2} 
dB_{\perp}/dx$ can be straightforwardly obtained from the flux 
density gradient.\cite{Brandt96}

\begin{figure}[tb]
\includegraphics[width=0.45\textwidth]{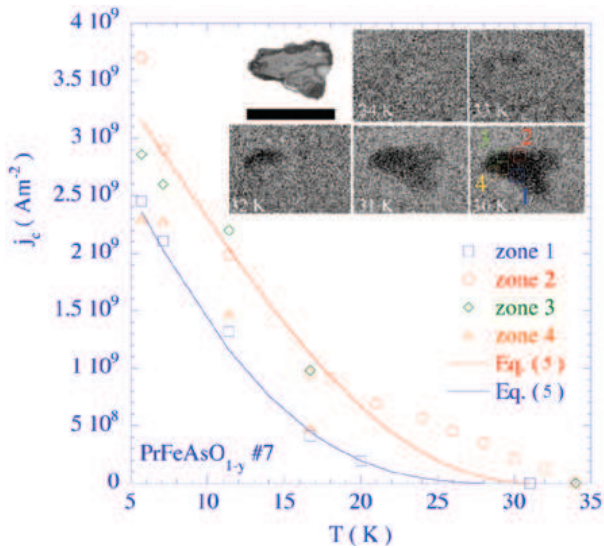}
\caption{(color online) Critical current density $j_{c}$ of PrFeAsO$_{1-y}$ crystal 7 at $H_{a} = 300$ Oe, 
as function of temperature. $j_{c}$ was 
determined from the slope of the local magnetic flux density, measured in the different regions of the 
crystal indicated in the Inset. The lower drawn line is a fit to 
Eq.~(\protect\ref{eq:1DCC-jc}), with $n_{d} = 1.5\times 10^{27}$ 
m$^{-3}$, the upper drawn line shows that the low temperature 
$j_{c}(T)$ in the strongly pinning regions is the same. Inset: Polarized light image of the 
crystal, and five DMO images of screening of $\Delta H_{a} =  
1$ Oe at $T = 34 - 30$ K (in steps of $- 1$ K). The regions 1 -- 4 over 
which $j_{c}$ was determined are indicated by drawn squares. }
\label{fig:jc-inhomogeneity}
\end{figure}

\begin{figure}[t]
\includegraphics[width=0.45\textwidth]{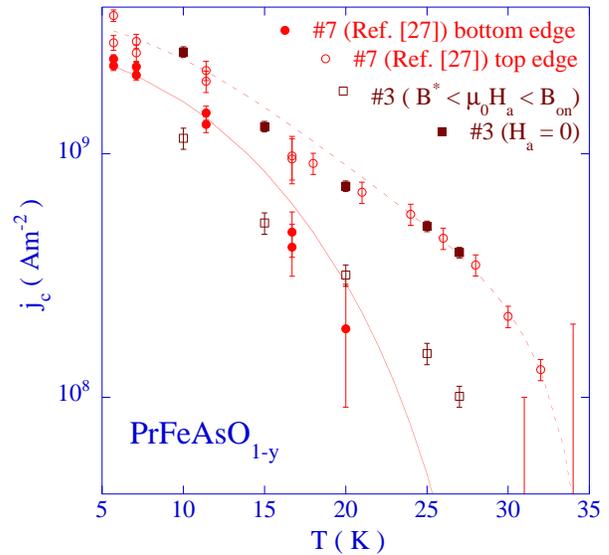}
\caption{(color online) PrFeAsO$_{1-y}$: critical current versus temperature, 
for two different crystals. Open and closed circles represent the upper 
and lower bounds of the $j_{c}$ of crystal \#7,\protect\cite{Okazaki2009}
the closed squares represent the zero-field $j_{c}$ of crystal \#3, 
while the open squares depict the magnitude of the critical current 
in the field regime between $B^{*}$, above which the strong pinning 
contribution becomes negligible, but below the onset field of the 
``fishtail'' effect, $B_{on}$. This value of $j_{c}$ is attributed to 
weak collective pinning by oxygen vacancies in the single vortex 
limit; the drawn line represents a fit to 
Eq.~(\protect\ref{eq:1DCC-jc}), with $n_{d} = 1.5\times 10^{27}$ m$^{-3}$.
The dashed line is a fit to  Eq.~(\protect\ref{eq:together}) summing strong 
and weak pinning contributions to the critical current at zero field, with 
parameter values $n_{d} = 1.5\times 10^{27}$ m$^{-3}$, $n_{i} = 
1 \times 10^{21}$ m$^{-3}$, and $f_{p,s} = 0.1 \varepsilon_{0}$}.
\label{fig:PrFeAsO-jc(T)}
\end{figure}

Resulting values of the critical current density in four areas of  
PrFeAsO$_{1-y}$ crystal \# 7 are shown in 
Fig.~\ref{fig:jc-inhomogeneity}, as function of temperature. The 
inset to the Figure reveals the inhomogeneity of $T_{c}$ for this 
particular crystal; the regions in which $j_{c}$ was measured are 
also indicated. It is found that $j_{c} = 3\pm 1 \times 10^{9}$ 
Am$^{-2}$ at the lowest measured temperature. The temperature 
dependence $j_{c}(T)$ depends on location. Low $j_{c}$ areas show a 
smooth decrease with temperature, whereas regions where $j_{c}$ is 
higher feature a crossover in the temperature dependence. 
Similar behavior is found in all investigated PrFeAsO$_{1-y}$ crystals, see Fig.~\ref{fig:PrFeAsO-jc(T)}. 
We shall, in section~\ref{sec:discussion}, attribute this behavior to 
the additive effect of weak collective pinning by oxygen dopant atoms, 
yielding a strong temperature dependence, and strong pinning, with a weak 
temperature dependence, coming from disorder of the doping level on 
the scale of 10 - 100 nm .

\begin{figure}[b]
\includegraphics[width=0.45\textwidth]{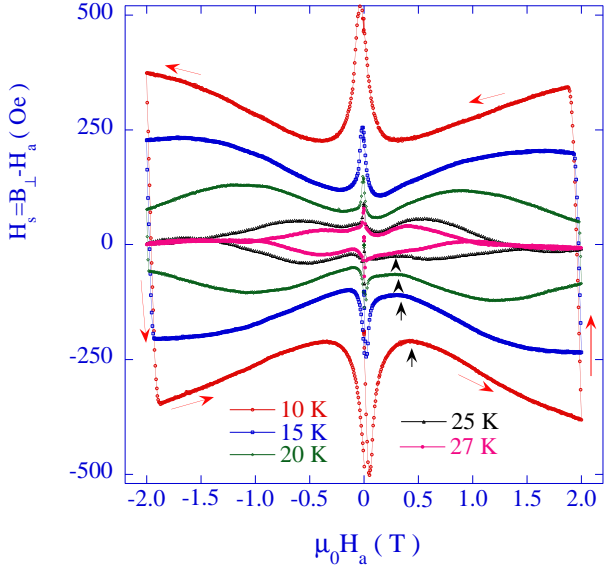}
\caption{(color online) Hysteresis loops ofthe ``self-field'', defined as $H_{s} = 
B_{\perp}/\mu_{0} - H_{a}$, as measured with a microscopic Hall sensor in the center of PrFeAsO$_{1-y}$ 
crystal \# 7, at the indicated temperatures. The values of the onset 
field $H_{on}$ are denoted by the vertical black arrows.}
\label{fig:PrFeAsO-fishtail}
\end{figure}

Measurements in higher magnetic fields were performed using the Hall 
array magnetometry technique. Typical results for the 
self-field, defined as $H_{s} = B_{\perp}/\mu_{0} - H_{a}$, measured 
on the central part of the top surface of crystal \# 7, are shown 
in Fig.~\ref{fig:PrFeAsO-fishtail}. The screening current density is 
proportional to the difference $\Delta H_{s}$ measured on the 
decreasing-- and increasing field branches, respectively. A clearly 
non-monotonous field-dependence of the critical current is observed, 
with the sustainable current density $j$ rapidly decreasing as the $H_{a}$ is first increased, 
followed by an intermediate regime of constant $j$. 
Fig.~\ref{fig:strong-pinning}(a) shows that the low-field behavior, a 
plateau up to $B^{*}$, followed by a power-law decrease $\sim B^{-5/8}$, is 
archetypal for a strong pinning contribution to the critical current. 
However, at intermediate fields, around 0.1 T in 
Fig.~\ref{fig:strong-pinning}(a), $j_{c}$ does not vanish, but 
saturates at a value $j_{c}^{SV} \sim 2$ - $3\times 10^{9}$ Am$^{-2}$ at low 
temperature. The temperature dependence of the zero-field-- and 
intermediate (constant) values of the critical current
are plotted in Fig.~\ref{fig:PrFeAsO-jc(T)}. One sees that the 
$j_{c}^{SV}$ contribution is spatially rather more homogeneous, 
and also that it  corresponds to the critical current measured in the most 
weakly pinning areas of the crystals. Below, we shall 
attribute this contribution to weak collective pinning by dopant 
atoms. The strong pinning contribution 
$j_{c}(0)$ strongly depends on the location at which it is measured, and 
it is responsible for the larger 
measured critical current densities. 

Finally, we turn to higher applied magnetic fields. It is observed 
that the hysteresis loops open up at a field $B_{on}$, corresponding to the 
increase of $j$ at the so-called ``fishtail'' or 
peak--effect.\cite{Prozorov2008,Kokkaliaris99,Paltiel,Klein2010,Shen2010,Yang2008,Yang2008ii,Wang2008} The $B_{on}(T)$ data are 
collected, together with the irreversibility fields determined from 
the appearance of a third harmonic component in the ac-response 
\cite{vdBeek95ii}, in Fig.~\ref{fig:phase-diagram}.

As in previous studies on other iron pnictide 
superconductors,\cite{Prozorov2008,Yang2008,Shen2010,Yang2008ii,Wang2008} 
the local flux density in Tesla fields is observed to decay with time, with a 
typical relaxation rate $S = (d \ln B_{\perp}/d \ln t) \sim -0.05$ for 
fields below $H_{on}$ and $S \sim -0.03$ for $H_{a} > H_{on}$. 
As in other studies,\cite{Shen2010} magnetic relaxation was not observed 
to affect the low-field MO data. it therefore does not affect 
the measured temperature dependence of the critical current density in 
what follows.

\subsection{NdFeAsO$_{1-x}$F$_{x}$}

\begin{figure}
    \includegraphics[width=0.45\textwidth]{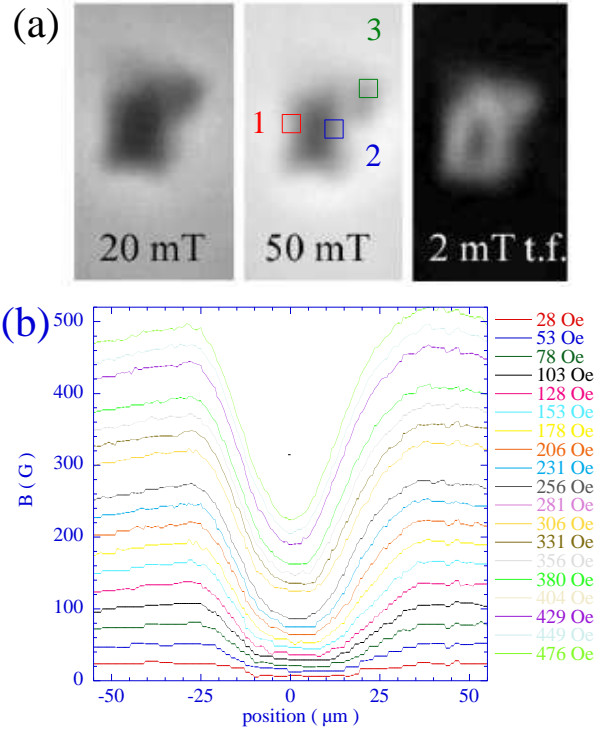}
    \caption{ (color online) (a) MOI flux penetration into  
    NdFeAsO$_{0.9}$F$_{0.1}$ crystal \# 1, at $T = 9.1$ K. The three 
    frames depict the flux density distribution after the application 
    of an applied field of 20 and 50 mT, and the remanent flux after 
    removal of the 50 mT applied field. (b) Flux density profiles 
    measured across the central part of the crystal.}
    \label{fig:NdFeAsO-profiles}
    \end{figure}

\begin{figure}
    \includegraphics[width=0.45\textwidth]{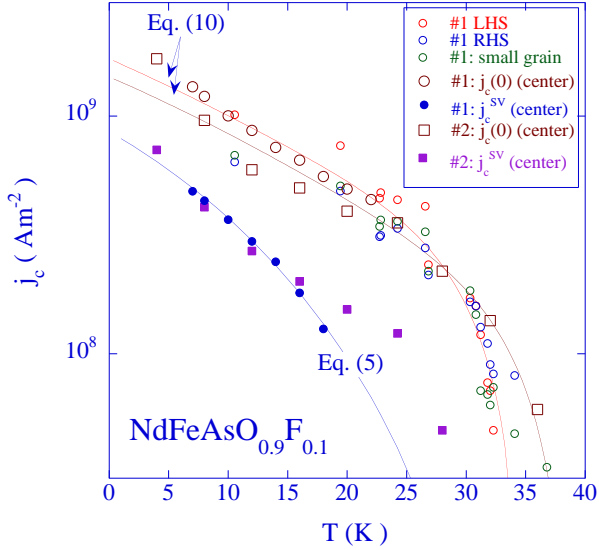}
    \caption{(color online) Local values of the critical current density $j_{c}$ in 
    the three regions of the NdFeAsO$_{0.9}$F$_{0.1}$ crystal \# 1, 
    outlined in the central panel of \protect\ref{fig:NdFeAsO-jc(T)}a. 
    Also shown are the values of $j_{c}^{SV}$ and $j_{c}^{s}(0)$ determined 
    for both investigated crystals. The first are compared to 
    Eq.~(\protect\ref{eq:1DCC-jc}) using $n_{d} = 1.5\times 10^{27}$ 
    m$^{-3}$ and  $D_{v} = 0.9$ nm (lower drawn line), while the 
    latter are fit to Eq.~(\protect\ref{eq:together}) with respective parameter 
    sets ($T_{c} = 37$ K, $n_{i} = 6 \times 10^{21}$ m$^{-3}$, $f_{p,s} = 0.1 \varepsilon_{0}$) and ($T_{c} = 
    35$ K, $n_{i} = 2\times 10^{22}$ m$^{-3}$, $f_{p,s} = 0.1 \varepsilon_{0}$). }
    \label{fig:NdFeAsO-jc(T)}
    \end{figure}

Fig.~\ref{fig:NdFeAsO-profiles}a shows magneto-optical images of flux 
penetration into NdFeAsO$_{0.9}$F$_{0.1}$ crystal \# 1. The 
sample turns out to be a bicrystal, with a similar spread in $T_{c}$ 
as observed in PrFeAsO$_{1-y}$. As shown by 
Fig.~\ref{fig:NdFeAsO-profiles}b,  flux distributions inside the 
crystalline grains are well-described by the Bean critical state 
model.\cite{Brandt96} Local values of the critical current density at 
$B_{\perp} \approx 300$ Oe were obtained in the same 
manner as described above. Results for the three regions outlined in 
the center panel of Fig.~\ref{fig:NdFeAsO-profiles}a are rendered as 
function of temperature in Fig.~\ref{fig:NdFeAsO-jc(T)}, together with 
results obtained by Hall probe magnetometry over the central regions 
of crystals \#1 and 2. Field-dependent results are shown in 
fig.~\ref{fig:strong-pinning}(b). The overall behavior 
recalls that reported in Ref.~\onlinecite{Wang2008}, and 
is very similar to that observed in PrFeAsO$_{1-y}$: high critical current areas correspond 
to a large local contribution of strong pinning, whereas the lower 
$j_{c}^{SV}$ measured at intermediate fields much above $B^{*}$ corresponds 
to the critical current density in the more weakly pinning areas of 
the crystals. In contrast to PrFeAsO$_{1-y}$, the strong pinning 
contribution outweighs $j_{c}^{SV}$ by a factor 2--3. NdFeAsO(O,F) 
crystal \# 2 shows a clear ``fishtail'' or peak-effect, the 
corresponding $B_{on}(T)$ values are plotted in 
Fig.~\ref{fig:phase-diagram}. A hint of a 
peak-effect is also observed in crystal \# 1, but the relative increase of the 
sustainable current density is much more modest than in the other 
investigated samples, 
with data resembling those of Ref.~\onlinecite{Wang2008}.
Finally, Fig.~\ref{fig:phase-diagram} shows that 
the irreversibility field measured from the onset of screening 
\cite{Kacmarcik2009} coincides with that determined from the onset of 
a third harmonic response in ac Hall-probe array magnetometry. 
Moreover, the irreversibility field $B_{irr}(T)$ for NdFeAs(O,F) and 
PrFeAsO$_{1-y}$ crystals with the same $T_{c}$ are, within 
experimental accuracy, identical.

\section{Discussion}
 \label{sec:discussion}
 

\subsection{Weak collective pinning}
We start by analyzing the critical current contribution $j_{c}^{SV}$ in terms of 
the weak collective pinning theory.\cite{Larkin79,Blatter94,Blatter2008} 
The vortex lattice order is characterized by the transverse and 
longitudinal displacement correlation lengths 
\begin{eqnarray}
	\langle | \mathbf u(R_{c},z) - \mathbf  u(0,z) |^{2} \rangle = r_{p}^{2}
	\label{eq:Rc} \\
	\langle | \mathbf u(\mathbf r,L_{c}) - \mathbf u(\mathbf r,0) |^{2} \rangle = 	r_{p}^{2}.
\end{eqnarray}
where $\mathbf u(\mathbf r , z)$ denotes the deformation field of the vortex 
lattice at position $(\mathbf r , z)$ (with $z \parallel \mathbf B$), and 
$r_{p} \sim \xi$ is the range of 
the pinning potential.\cite{Brandt86} The transverse displacment correlation 
length 
 \begin{equation}
    R_{c} = \left(\frac{\varepsilon_{0}\xi}{2\Phi_{0}j_{c}}\right)^{1/2}
    \label{eq:3DCC-Rc}
\end{equation}
can be obtained, without a priori assumptions, from the value of the 
critical current density. Using the appropriate parameters 
(Table~\ref{table:parameters}), one has, for $j_{c}^{SV}( 5 \, \mathrm K ) = 
3 \times 10^{9}$ Am$^{-2}$, $R_{c} = 40$~nm in single crystalline 
PrFeAsO$_{1-y}$, and $R_{c} = 56$ nm corresponding to $j_{c}^{SV}( 5 \,
\mathrm K) \sim 1 \times 10^{9}$ Am$^{-2}$ in 
NdFeAsO$_{0.9}$F$_{0.1}$. These values are much smaller than the 
intervortex spacing at 300 Oe, at which the data in 
Figs.~\ref{fig:jc-inhomogeneity}, \ref{fig:PrFeAsO-jc(T)}, and 
\ref{fig:NdFeAsO-jc(T)} were obtained. The pinning-induced 
displacement of each vortex is thus independent of that of neighboring 
vortices. In this so-called single vortex pinning limit, one may now 
estimate the longitudinal displacement correlation length as
\begin{equation}
    L_{c} = \xi    
    \left(\frac{\sqrt{3}\varepsilon_{\lambda}^{2}\varepsilon_{0}}{2j_{c}\Phi_{0}\xi}\right)^{1/2};
    \label{eq:1DCC-Lc}
\end{equation}
one finds $L_{c} \approx 20$ nm and 10 nm for PrFeAsO$_{1-y}$ and 
NdFeAsO$_{0.9}$F$_{0.1}$, respectively. This length largely exceeds 
the spacing of the FeAs planes, which clearly establishes pinning as 
being in the three-dimensional single-vortex (3DSV) 
limit.\cite{Blatter94,Blatter2008} 

From here on, we show that the critical current density in the (1111) 
iron oxypnictide superconductors can be understood as arising from 
mean-free path variations induced by the dopant atoms, oxygen 
vacancies in the case of PrFeAsO$_{1-y}$, and F ions in the case of 
NdFeAsO$_{0.9}$F$_{0.1}$. The  pinning force of a single defect is 
expressed as $f_{p} \sim 0.3 
g(\rho_{D}) 
\varepsilon_{0} \left( \sigma_{tr}/\pi 
\xi^{2}\right) \left( \xi_{0}/\xi \right) $, where $\sigma_{tr} = \pi D_{v}^{2}$ is the 
transport scattering cross-section, $D_{v}$ is the effective ion 
radius, and $g(\rho_{D})$ is the Gor'kov function. The disorder 
parameter $
\rho_{D} 
= \hbar v_{F}/2\pi T_{c} l\sim \xi_{0}/l$, with 
$v_{F}$ the Fermi velocity, $l$ the mean free path, and $\xi_{0} \approx 1.35 \xi(0)$ the (temperature-independent) 
Bardeen-Cooper-Schrieffer coherence length.\cite{Thuneberg84,Blatter94} The critical current is determined by the fluctuation of the 
elementary pinning force, $\langle f_{p}^{2} \rangle$, and reads 
\cite{vdBeek2002}
\begin{eqnarray}
    j_{c}^{SV} & \approx & j_{0}   
    \left[ \frac{0.1 n_{d} D_{v}^{4}}{\varepsilon_{\lambda} \xi} 
    \left( \frac{\xi_{0}}{\xi} \right)^{2} \right]^{2/3} 
    \label{eq:1DCC-jc}
    \\
    & \propto &   \left[\frac{\lambda(0)}{\lambda(T)}\right]^{2} 
    \left( 1 - \frac{T}{T_{c}} \right)^{\alpha}.
\end{eqnarray}

The numerical factor under the parentheses in Eq.~(\ref{eq:1DCC-jc}) 
depends on the precise type of scattering.\cite{Thuneberg84} 
Since the temperature dependences $\lambda(0)/\lambda(T)$ and 
$\varepsilon_{\lambda}(T)$ are known from 
Refs.~\onlinecite{Okazaki2009} and 
\onlinecite{Pribulova2009} (yielding $\alpha \sim 2$ for  
PrFeAsO$_{1-y}$ and $\alpha \sim 1.5$ for NdFeAsO$_{0.9}$F$_{0.1}$), one is in the position where a full 
consistency check of both the magnitude and the temperature dependence 
of $j_{c}$ is possible.\cite{T-dependence} 

In the case of PrFeAsO$_{1-y}$, Eq.~(\ref{eq:1DCC-jc}), we start from 
the hypothesis that O vacancies are responsible for the lion's share 
of flux pinning. The 
ion radius $D_{v} = 1.46\times 10^{10}$ m. Inserting this value into 
Eq.~\ref{eq:1DCC-jc} reproduces the low-temperature value $j_{c}^{SV} = 
3\times 10^{9}$ Am$^{-2}$ with the single free parameter $n_{d} 
\approx 1.5 \times 10^{27}$ m$^{-3}$. This nicely corresponds to 0.1 O 
vacancy per formula  unit (half a unit cell of volume 65 \AA$^{3}$). 
Eq.~(\ref{eq:1DCC-jc}) reproduces the 
low-$T$ temperature dependence of the critical current density in the 
high-$j_{c}$ regions, and the temperature dependence over the 
full range from 5 K to $T_{c}$ in the low-$j_{c}$ regions. The 
spatially more homogeneous contribution to the critical current 
density of oxygen deficient single crystalline PrFeAsO$_{1-y}$ is therefore well-described by 
pinning by O-vacancies by the $\delta \kappa$ mechanism.

In the case of NdFeAsO$_{0.9}$F$_{0.1}$, the analysis is 
hindered by our ignorance of the effective scattering cross-section: 
doping is through chemical substitution, not oxygen depletion. 
If one adopts the view that F substitution is at the origin of 
pinning, one has $n_{d} \sim 1.5 \times 10^{27}$ m$^{-3}$ for our 
average doping level. To reproduce the value of the 
measured low-$T$ $j_{c} \approx 1\times 10^{9}$ Am$^{-2}$ then requires 
$\sigma_{tr} = 1.5\times 10^{-20}$ m$^{2}$, corresponding to an effective defect radius of 
0.9 \AA (this can be compared to the F ion radius of 1.3 \AA). 
The temperature dependence of $j_{c}^{SV}$ is  again very well described by 
Eq.~(\ref{eq:1DCC-jc}). It is not quite as strong as in 
PrFeAsO$_{1-y}$, an effect that can be attributed to the different 
$T$--dependence of the penetration depth [$\lambda(T)$ nearly perfectly 
follows $ \lambda^{-2} \sim (1-t^{2})$] and of the anisotropy 
ratio ($\varepsilon_{\lambda}$ seems to be nearly independent of 
temperature in NdFeAs$_{0.9}$F$_{0.1}$).\cite{Pribulova2009}

\begin{figure}[tb]
\includegraphics[width=0.45\textwidth]{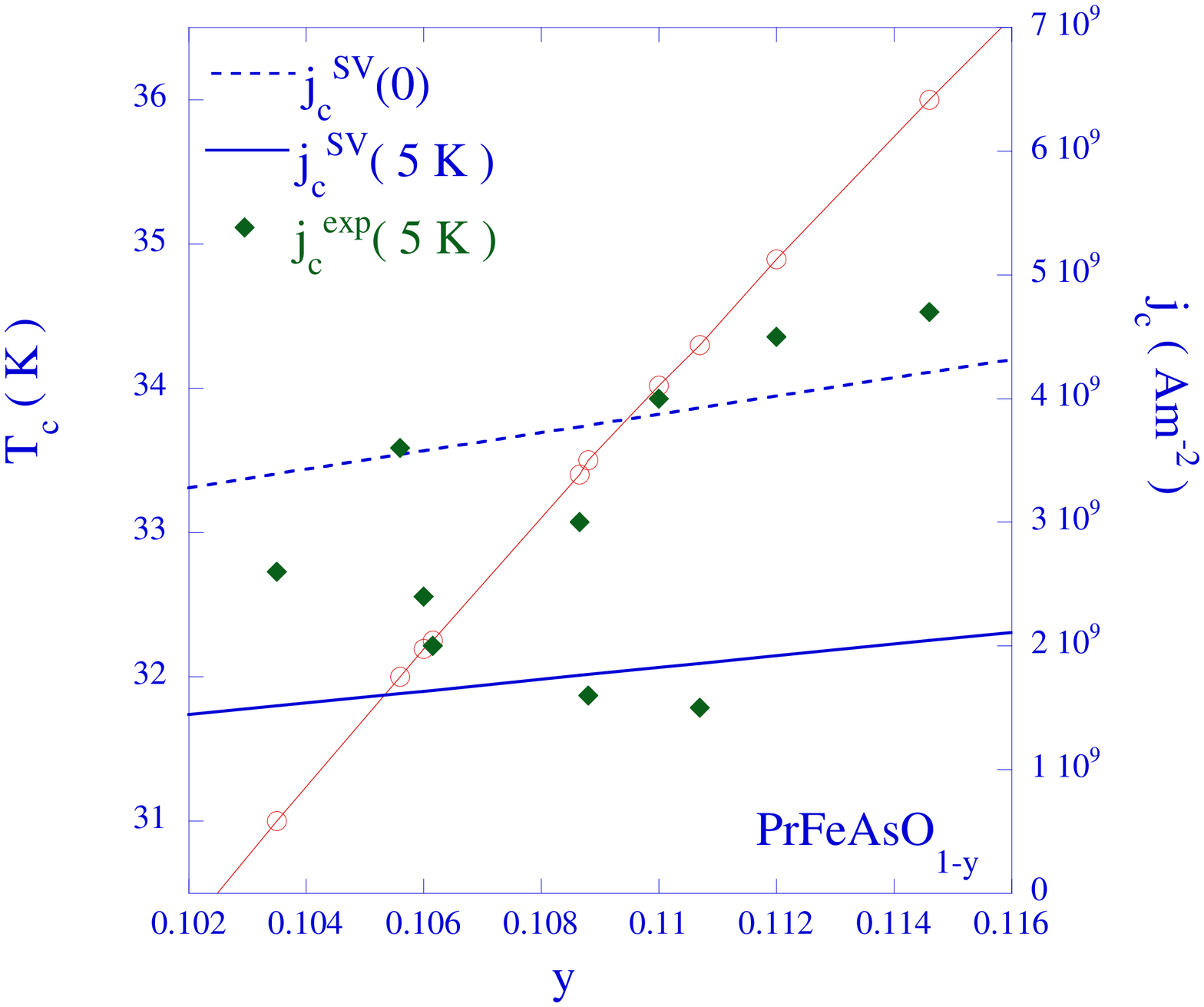}
\caption{(color online) PrFeAsO$_{1-y}$: critical current versus doping level, as 
determined from the phase diagram of 
Ref.~\protect\onlinecite{Rotundu2009}. Open 
circles indicate the critical temperature, filled diamonds indicate 
the critical current density as measured in various locations in the four 
different crystals of Fig.~\protect\ref{fig:PrFeAsO-jc(T)}. The 
dashed and drawn lines indicate the critical current density expected 
from weak collective pinning by oxygen vacancies, 
Eq.~(\protect\ref{eq:1DCC-jc}), at $T = 0$ and $T = 5$ K, 
respectively (here we suppose that $\lambda_{ab}^{-2} \propto 
T_{c}$, as in Ref.~\protect\onlinecite{Uemura89}.}
\label{jc-doping}
\end{figure}

\subsection{Spatial variations of $j_{c}$ and link with doping}

Both investigated (1111) compounds show spatial variations of both the 
critical temperature $T_{c}$ and the low-temperature critical current density $j_{c}$. 
It is tempting to correlate the two: knowing the temperature 
dependence of both the superfluid density $\lambda^{-2}$ and the 
anisotropy ratio, as well as the evolution of the respective $T_{c}$ 
vs. doping phase diagrams\cite{Rotundu2009,Chen2008}, 
Eq.~(\ref{eq:1DCC-jc}) predicts what the dependence $j_{c}(T_{c})$ should 
be. In the case of PrFeAsO$_{1-y}$, our measurements yield sufficient 
statistics for the expected increase of $j_{c}$ 
with $T_{c}$ to be, indeed, observed. In the considered portion of the phase 
diagram, the more vacancies are added, the higher $T_{c}$, but also 
the stronger the pinning. Fig.~\ref{jc-doping} shows a compilation of 
critical temperatures and low-temperature critical currents of all
investigated  regions in all our PrFeAsO$_{1-y}$ crystals. The 
experimental data follow the dependence of the low temperature $j_{c}$ as this 
follows from Eq.~(\ref{eq:1DCC-jc}), even though this dependence is 
weak. The contribution to this dependence via $n_{d}$, arising from the 
addition of oxygen vacancies, is actually weaker than the expected 
contribution from the doping dependence of the 
the superfluid density, which we have assumed to follow the 
relation $\lambda^{-2} \propto T_{c}$.\cite{Uemura89,Drew2008} 
Significant scatter due to the strong pinning contribution remains in 
Fig.~\ref{jc-doping}, 
which we shall attribute to the presence of 
doping inhomogeneity on the 10 -- 100 nm scale .

In the framework of weak collective pinning, the  observed spatial variation of $j_{c}$ in NdFeAs$_{0.9}$F$_{0.1}$ 
would, if attributed to the macroscopic variation of the dopant atom 
density , correspond to a variation of the doping level of $x = 0.1 \pm 0.03$, 
within a given single crystal. The concomittant $T_{c}$ variation 
would be from 26 K to nearly 50 K, which is not what is observed by 
DMO. Moreover, and contrary to the observation in PrFeAsO$_{1-y}$, the critical 
current density of the investigated crystal is larger in areas with low 
$T_{c}$, both as far as different regions of crystal \# 1 are 
concerned, as the observed differences between crystals \# 1 and 2.
In the absence of sufficient statistics, we tentatively 
ascribe this behavior to the presence of strong background pinning.

\begin{figure}[tb]
\includegraphics[width=0.45\textwidth]{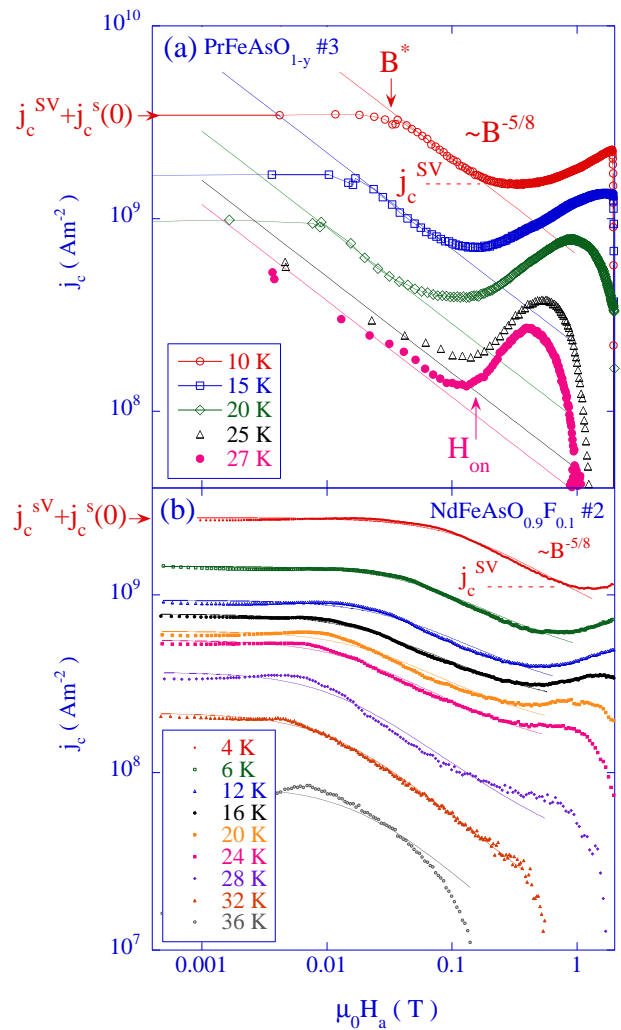}
\caption{(color online) (a) PrFeAsO$_{1-y}$: Double logarithmic plot of the critical current 
versus magnetic field. The drawn lines show the power $B^{-5/8}$ 
expected from the strong pinning contribution. (b) \em ibid \rm, for 
NdFeAsO$_{0.9}$F$_{0.1}$. Drawn lines show model fits to 
Eq.~\protect\ref{eq:together}.}
\label{fig:strong-pinning}
\end{figure}

\subsection{Strong pinning background}
\label{section:strong}

As described in section~\ref{section:results}, the spatial variation 
of the critical current density in single crystalline PrFeAsO$_{1-y}$ 
is reflected in the temperature dependence of $j_{c}$, higher local 
$j_{c}$ corresponding to the presence of a break in the temperature dependence. 
Also, the higher local critical current densities are responsible 
for the low-field $j_{c}$--peak observed in 
Fig.~\ref{fig:PrFeAsO-fishtail}, which cannot be explained within the 
single-vortex collective pinning framework. There 
must therefore be supplementary sources of pinning, inhomogeneously distributed 
throughout the samples, with a temperature dependence that is weaker than that 
of the weak collective pinning described above. 

The field dependence of the associated critical current density, 
a plateau, followed by a power-law decrease $j_{c} \propto B^{-5/8}$,
is in very satisfactory agreement with the theory of strong pinning 
developped in Refs.~\onlinecite{Ovchinnikov91,vdBeek2002}. 
In the presence of a density $n_{i}$ of strong pins of size larger 
than the coherence length, one has\cite{vdBeek2002}
\begin{eqnarray}
    j_{c}^{s}(0) & = & \frac{\pi^{1/2} n_{i}^{1/2} j_{0}}{\varepsilon_{\lambda}} 
    \left( \frac{f_{p,s}\xi_{ab}}{\varepsilon_{0}}\right)^{3/2} 
    \hspace{1mm} \left( B < B^{*} \right) 
    \label{eq:1D-strong} \\
    j_{c}^{s}(B) & \approx & \frac{2 n_{i} 
    j_{0}}{\varepsilon_{\lambda}^{5/4}\xi_{ab}^{1/2}} 
    \left( \frac{f_{p,s}\xi_{ab}}{\varepsilon_{0}} \right)^{9/4} \left( 
    \frac{\Phi_{0}}{B}\right)^{5/8} \nonumber \\
     & & 
    \hspace{37mm} \left( B > B^{*} \right).
    \label{eq:3D-strong}
\end{eqnarray}
The crossover field $B^{*} = 0.74 \varepsilon_{\lambda}^{-2} \Phi_{0} \left( n_{i}/\xi_{ab} 
\right)^{4/5} \left( f_{p,s}\xi_{ab}/\varepsilon_{0} 
\right)^{6/5}$ is determined as that above which the so-called vortex trapping area of a 
single pin is limited by intervortex interactions.\cite{vdBeek2002} 
The identification of the experimental $j_{c}(0)$ with 
Eq.~(\ref{eq:1D-strong}), and of the power-law decrease with 
Eq.~(\ref{eq:3D-strong}), allows for the determination of the elementary 
pinning force $f_{p,s}$ of a strong pin from the ratio $ \left[ dj_{c}^{s}(B) / 
dB^{-5/8}\right] / \left[j_{c}^{s}(0)\right]^{-2}$. It is found that 
$f_{p,s}(0) = 2\times 10^{-13}$ N for both investigated 
compounds, with a temperature dependence coinciding with that of the 
superfluid density. Hence, we find a measured $f_{p,s} \sim 0.1 
\varepsilon_{0}$. The density of strong pins can be 
straightforwardly estimated from $B^{*}$: $n_{i} \approx 1 \times 10^{21}$ m$^{-3}$ 
for PrFeAsO$_{1-y}$, and  $n_{i} \approx 6 \times 10^{21}$ m$^{-3}$ and $\approx 2 \times 
10^{22}$ m$^{-3}$ for NdFeAsO$_{0.9}$F$_{0.1}$ crystals \# 1 and 2, respectively.

These data can be compared to the results of TEM observations. The first 
candidate strong pins are extended (nm-sized) pointlike inclusions or 
precipitates, such as observed in Fig.~\ref{fig:TEM}b.
Assuming such defects to be non-superconducting, one would have 
$f_{p,s} \sim  \varepsilon_{0} \left(  D_{i} / 4 \xi_{ab} \right) 
\ln \left( 1 + D_{i}^{2}/2 \xi_{ab}^{2} 
\right)$. Typical observed defect dimensions are $D_{i} \approx 2$ -- 5 
nm, yielding $f_{p,s} \sim 0.1$ -- $1.1 \varepsilon_{0}$ at 
low temperature. Therefore, the smaller defects of radius 2 nm might 
do the job, were it not that the temperature dependence expected for 
such voids is at odds with experiment.

Next, the observed undulations of the FeAs layers impose an intermittant bending of 
vortex lines as these move through the crystal lattice. The necessary 
force to produce this bending can be estimated as the product of the 
line tension $\varepsilon_{\lambda}^{2}\varepsilon_{0} $ and the 
variance $(\delta \alpha)^{2}$ of the tilt angle; here, $\alpha$ 
corresponds to the buckling angle. Such a mechanism 
would yield the experimental temperature dependence of 
$f_{p,s}$, but, at $10^{-4}\varepsilon_{0}$, grossly underestimates the measured elementary 
force. 

Third, the higher strong pinning critical current density observed for 
lower doped NdFeAs(O,F) could be linked to the observation of phase 
coexistence in the underdoped state of this material.\cite{Drew2008} 
Without going as far as invoking the presence of nm-scale magnetically ordered 
regions in our crystals, the idea of phase coexistence suggests that 
there are spatial fluctuations of the dopant atom density on the 
scale of several nm. The ensuing dispersion of weakly superconducting regions with critical 
temperature $T_{c}-\delta T_{c}$ inside a more strongly superconducting matrix would certainly 
lead to flux pinning. Its description would be similar to that of 
non-superconducting precipitates, but with a smaller pinning energy, 
a vortex passing through an area of lower $T_{c}$ gaining only a 
fraction $\delta T_{c}/T_{c}$ of the condensation 
energy $\varepsilon_{0}/4 \xi^{2}$. Assuming the condensation energy to be proportional to the 
critical temperature, the pinning force can be written as
\begin{eqnarray}
  f_{p,s} & \approx &  \left[  \varepsilon_{0}\left( t \right)  - 
  \left( 1 - \frac{\delta T_{c}}{T_{c}} \right) \varepsilon_{0} 
  \left( \tilde{t}  \right) \right] \nonumber \\
 &  & \hspace{1cm} \times    \left(  \frac{D_{i}}{4 \xi_{ab}} \right) \ln \left( 1 + \frac{D_{i}^{2}}{2 \xi_{ab}^{2}} \right).
\label{eq:Tc-disorder}
\end{eqnarray}
with $t \equiv T/T_{c}$, and $\tilde{t} \equiv T/\left(T_{c}-\delta 
T_{c}\right)$.
For small spatial variations of the critical temperature, \em e.g. \rm 
$\delta T_{c}/T_{c} \sim 0.05$ or $\delta T_{c} \approx 1.5$ K, and $D_{i} \sim 5 - 10$ nm, Eq.~(\ref{eq:Tc-disorder}) nicely 
mimics the measured temperature dependence $f_{p,s}(T) \sim 
\varepsilon_{0}(T)$. As shown in Figs.~\ref{fig:PrFeAsO-jc(T)}, 
\ref{fig:NdFeAsO-jc(T)}, and \ref{fig:strong-pinning}(b), the total critical current density, obtained by 
summing Eqs.~(\ref{eq:1D-strong}) [with (\ref{eq:Tc-disorder})
inserted] and (\ref{eq:1DCC-jc}),
\begin{equation}
    j_{c} =  j_{c}^{SV} + j_{c}^{s}.
    \label{eq:together}
\end{equation}
is also in good agreement with experimental observations. One is thus 
lead to the conclusion that, in addition to the macroscopic 
inhomogeneity of doping level, there also exists an inhomogeneity on 
the nano-scale, much similar to that reported by Yamamoto {\em et al.} 
in Ba(Fe$_{0.9}$Co$_{0.1}$)$_{2}$As$_{2}$. However, the doping level 
modulation, necessarily of the order of the $T_{c}$--variation, $\delta 
T_{c}/T_{c} \sim 0.05$, that 
explains the strong pinning contribution, is far too small 
to support any claims of phase coexistence in the underdoped (1111) 
pnictides investigated here. If similar disorder should exist for smaller doping levels, near the 
superconductivity onset, one would have $\delta T_{c} \sim T_{c}$, and 
a near certain coexistance of magnetic and superconducting regions. 
This is a premise that needs further investigation.

For completeness, one may also contemplate surface roughness  
as a source of flux pinning.\cite{Mathieu88,Simon95,Lazard2002} The critical current density is then 
determined by the force needed to push a vortex line out of a surface trough 
or across a ridge, and reads, in the limit of small magnetic 
fields\cite{vdBeek2002,Flippen95}
\begin{equation}
    j_{c}^{TV} =  \frac{\pi \varepsilon_{0}}{\Phi_{0} d} \frac{\delta 
    d}{D} \hspace{1cm} \left( B \lesssim \frac{\Phi_{0}}{D^{2}} \right).
    \label{eq:TV}
\end{equation}
Here, $d$ would be the crystal thickness, $D$ the spacing between surface 
defects or troughs, and $\delta d$ the typical ridge height, that is, 
the variance of the thickness. In 
Refs.~\onlinecite{Mathieu88,Simon95,Lazard2002}, the ratio $\delta 
d/D = \sin \theta_{c}$ is interpreted as the sine of a ``contact angle'' $\theta_{c}$. 
In the Mathieu-Simon model\cite{Mathieu88,Simon95,Lazard2002} the 
field dependence is expected to correspond to that of the vortex chemical 
potential, \em i.e. \rm the equilibrium magnetization. This is not 
observed. Moreover, if one reinterprets the experimental $f_{p,s}$ and $n_{i} \sim 
2/dD^{2}$ in terms of surface pinning, one finds a ratio of ridge 
height to ledge width $\delta d/D \sim 2$, for a ledge separation 
of $\sim 20$ nm. Such a high aspect ratio would mean that the surface defects are  
located on the crystal edge, since the alternative, cracks on the 
surface, are not observed. Strong pinning by impurities, located in 
surface regions only, leads to the same dependences 
(\ref{eq:1D-strong},\ref{eq:3D-strong}), but with  
$3\times 10^{16}$ defects $m^{-2}$. 

\begin{figure}[tb]
\includegraphics[width=0.45\textwidth]{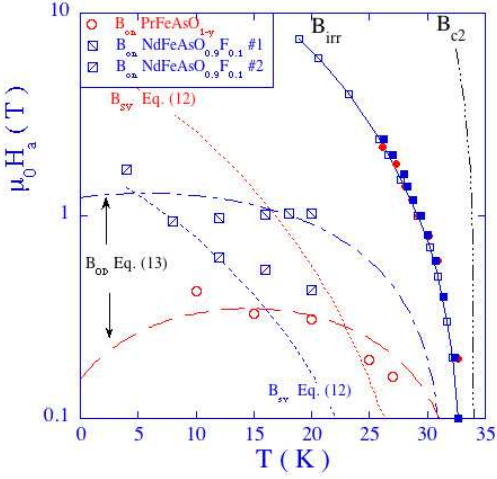}
\caption{
(color online) $(B, T)$ vortex matter phase diagram for (1111) iron 
pnictide superconductors. (Red) Circles indicate measurements on 
PrFeAsO$_{1-y}$, while (blue) squares show results on 
NdFeAsO$_{0.9}$F$_{0.1}$. Closed 
circles show the irreversibility field $B_{irr}(T)$ measured from the 
onset of a third harmonic reponse from ac Hall probe magnetometry; 
open (blue) squares show the screening onset data of 
Ref.~\protect\onlinecite{Kacmarcik2009}. Peak effect onset fields for 
both compounds are indicated by barred squares 
(NdFeAsO$_{0.9}$F$_{0.1}$) and open (red) circles (PrFeAsO$_{1-y}$). Dotted 
lines show the single-vortex to bundle pinning crossover described by 
Eq.~(\ref{eq:B_SV}), while dashed-dotted lines indicate the 
order-disorder field described by Eq.~(\ref{eq:mikitik}). 
}
\label{fig:phase-diagram}
\end{figure}

\subsection{Fishtail effect and phase diagram}

The knowledge of pinning parameters of the (1111) superconductors 
under study allows one to confront features of the mixed-state 
$(B,T)$-phase diagram with theoretical models. In particular, the 
fishtail effect at $B_{on}(T)$ was attributed to a crossover in vortex dynamics as, 
with increasing magnetic field, one leaves the single vortex pinning 
regime for the bundle pinning regime,\cite{KrusinElbaum92,Civale94} or the occurence 
of a first order phase transition from an ordered, ``elastically 
pinned'' low-field vortex phase, the so-called 
Bragg-glass,\cite{Giamarchi97} to a high field disordered phase 
characterized by the presence of topological 
defects.\cite{Kierfeld2000,Mikitik2001} The latter scenario has been 
unambiguously verified in the high temperature superconductors 
YBa$_{2}$Cu$_{3}$O$_{7-\delta}$ and 
Bi$_{2}$Sr$_{2}$CaCu$_{2}$O$_{8+\delta}$,\cite{Kokkaliaris99,Rassau99,vdBeek2000ii}, 
in the cubic superconductor (Ba,K)BiO$_{3}$,\cite{Klein2001} in 
NbSe$_{2}$,\cite{Paltiel} as well as in MgB$_{2}$.\cite{Klein2010} 

In the first case, the onset field $B_{on}$ should coincide with the 
single-vortex- to bundle pinning crossover field $B_{SV}$, determined by the 
equality of $R_{c}$ [see Eq.~(\ref{eq:3DCC-Rc})] and the vortex 
spacing $a_{0}$:
\begin{equation}
    B_{SV} \sim 40 B_{c2} \left( \frac{j_{c}^{SV}}{j_{0}} \right).
    \label{eq:B_SV}
\end{equation}
Inserting the experimentally obtained $j_{c}^{SV}$ into 
Eq.~(\ref{eq:B_SV}) yields the dotted lines in 
Fig.~\ref{fig:phase-diagram}. Clearly, while the experimental $B_{on}$ 
data for more strongly pinning PrFeAsO$_{1-y}$ lie below those for more weakly 
pinning NdFeAsO$_{0.9}$F$_{0.1}$, Eq.~\ref{eq:B_SV} predicts otherwise. 
Therefore, even if the peak effect onset lies in the vicinity of the 
single-vortex to bundle pinning crossover, it cannot be directly 
associated with it.

On the other hand, the vortex ensemble can undergo a structural 
transition whereby it lowers its energy by adapting 
itself more efficiently to the underlying pinning potentiel, at the 
expense of the generation of topological 
defects.\cite{Giamarchi97,Kierfeld2000,Mikitik2001} In the absence of 
a theory for this order-disorder transition of the vortex lattice, a 
Lindemann-like criterion was developped in 
Refs.~\onlinecite{Ertas94,Giamarchi97} and \onlinecite{Vinokur98} in order to, at least, estimate 
its position in the $(B,T)$--plane. The Lindemann approach considers 
that topological defects can be generated when pinning is 
sufficiently strong to provoke the wandering of vortex lines outside 
their confining cage formed by the nearest neighbor flux lines. The different 
results \cite{Ertas94,Vinokur98} have been summarized in Ref.~\onlinecite{Mikitik2001}. 
In the regime of single vortex pinning, relevant for collective 
pinning in the (1111) compounds, the position of the 
order-disorder transition is given by 
\begin{equation}
    A b_{SV}^{3/5}b_{OD}^{2/5}\left[ 1 + 
    \frac{F_{T}(t)}{b_{SV}^{1/2}\left( 1 - b_{OD} \right)^{3/2}} \right] = 
    2 \pi c_{L}^{2}
    \label{eq:mikitik} 
\end{equation}
where $b_{on} \equiv B_{OD}/B_{c2}$, $b_{SV} = B_{SV}/B_{c2}$, $c_{L} \sim 0.1$ 
is the Lindeman number, $A$ is a numerical constant, $t = t/T_{c}$, and 
$F_{T}(t) = 2 t \left( Gi / 1-t^{2} \right)^{1/2}$. The use of the 
parameters of Table~\ref{table:parameters}, the experimentally 
measured $j_{c}^{SV}$, and $A = 4$ yields the dashed lines in 
Fig.~\ref{fig:phase-diagram}. These show more than satisfactory 
agreement with the experimentally measured positions of $B_{on}$. We 
therefore conclude that, most likely, a bulk order-disorder 
transition of the vortex lattice lies at the origin of the peak 
effect in (1111) pnictide superconductors. However, more work, 
especially on vortex dynamics and possible hysteresis associated with 
the transition, should be performed to ascertain this.

\section{Conclusion}

It is found that superconducting iron pnictide  single crystals show 
significant spatial variations of both the critical temperature 
$T_{c}$ and the critical current density $j_{c}$. Variations of these 
quantities on the macroscopic scale, from several to several hundred $\mu$m, 
are at the origin of a smearing of globally measured properties, and notably 
of the width of the superconducting transition. This implies the 
necessity of local measurements, such as magneto-optical imaging or 
Hall-probe magnetometry, to extract superconducting parameters. From 
such local measurements, it is found that the critical current in iron 
oxypnictide superconductors of the (1111) family of compounds arises from 
two distinct contributions. The first is weak collective pinning by dopant 
atoms or vacancies, vortex lines being pinned by the small scale fluctuations 
of the local dopant atom density. The pinning mechanisme is identified as being
due to mean-free path variations in the vortex core ($\delta \kappa$ mechanism). 
This means that dopant atoms should also be effective quasi-particle scatterers. 
The second pinning contribution manifests itself at low fields. The corresponding  
critical current contribution can be completely parametrized by the strong pinning 
theory developped in Refs.~\onlinecite{Ovchinnikov91,vdBeek2002}, which means that 
extended defects are at its origin. An analysis of the magnitude and 
field-dependence of this strong pinning contribution shows that spatial variations of the doping level on the 
scale of several dozen to one hundred nm may be at stake. These 
variations do not support the possible coexistence of the 
anti-ferromagnetic metallic and the superconducitng phases. Finally, 
we contend that a bulk order-disorder transition of the vortex 
ensemble is at the origin of the 
``fishtail'' or peak effect
observed
in the critical current in sub-T fields. 

\section*{Acknowledgements} We wish to acknowledge V. Mosser for 
providing the Hall sensor arrays. This work was supported by KAKENHI from JSPS, and by
Grant-in-Aid for the Global COE program ``The Next Generation
of Physics, Spun from Universality and Emergence'' from MEXT, Japan. R.O. was supported by the JSPS Research Foundation for Young Scientists. Work at the Ames Laboratory was supported by the Department of Energy, Basic Energy Sciences under Contract No. DE-AC02-07CH11358.


\begin{thebibliography}{99}
\bibitem{T-dependence} The different temperature dependences arise from the different temperature dependence of $\varepsilon_{\lambda}$ in the two compounds; Whereas $\varepsilon_{\lambda}$ increases with temperature in PrFeAsO$_{1-y}$ it is nearly temperature independent in NdFeAs$_{0.9}$F$_{0.1}$.
\end{thebibliography}
\end{document}